\documentstyle[12pt,epsf]{article}
\newcommand{\be}{\begin{equation}}
\newcommand{\ee}{\end{equation}}
\newcommand{\bea}{\begin{eqnarray}}
\newcommand{\eea}{\end{eqnarray}}
\newcommand{\bdm}{\begin{displaymath}}
\newcommand{\edm}{\end{displaymath}}
\newcommand{\no}{\nonumber \\}

\newcommand{\qbar}{\overline{\rule[0.42em]{0.4em}{0em}}\hspace{-0.5em}q}

\newcommand{\Kbar}{\,\overline{\rule[0.75em]{0.7em}{0em}}\hspace{-0.85em}K}
\newcommand{\Lbar}{\,\overline{\rule[0.7em]{0.5em}{0em}}\hspace{-0.75em}{\cal
L}}
\newcommand{\eff}{{e\hspace{-0.1em}f\hspace{-0.18em}f}}
\newcommand{\QCD}{{\mbox{\scriptsize Q\hspace{-0.1em}CD}}}

\newcommand{\R}{{\scriptscriptstyle R}}
\renewcommand{\L}{{\scriptscriptstyle L}}
\newcommand{\al}{&\!\!\!\!}
\newcommand{\fs}{\; \; .}
\newcommand{\co}{\; \; ,}

\begin{document}

\begin{titlepage}
\begin{flushright}
$\rule{0cm}{0cm}$
\end{flushright}
\vspace*{2cm}
\begin{center}
{\LARGE {\bf Bounds on the light quark masses}}\\ \vspace{0.8cm}
H. Leutwyler\\Institut f\"{u}r theoretische Physik der Universit\"{a}t
Bern\\Sidlerstr. 5, CH-3012 Bern, Switzerland and\\
CERN, CH-1211 Geneva, Switzerland\\
\vspace{0.6cm}
January 1996\\
\vspace{3cm}
{\bf Abstract} \\
\vspace{1.2em}
\parbox{30em}{
The corrections to the current algebra mass formulae for the
pseudoscalar mesons are analyzed by means of a simultaneous exansion in
powers of the light quark masses and powers of $1/N_c$. The relative
magnitude of the two expansion parameters is related to the mass ratio
$M_\eta^2/M_{\eta^\prime}^2$, which represents a quantity of order $N_c m_s$.
A set of mass formulae is derived, including an inequality, which leads
to bounds for the ratios $m_u/m_d$ and $m_s/m_d$.

}\\ \vspace{3cm}
\rule{30em}{.02em}\\
{\footnotesize Work
supported in part by Schweizerischer Nationalfonds}
\end{center}
\end{titlepage}

The low energy properties of QCD are governed by an approximate,
spontaneously broken symmetry, which originates in the fact that three of the
quarks happen to be light. If $m_u,m_d,m_s$ are turned off, the symmetry
becomes exact. The spectrum of the theory then contains eight strictly massless
pseudoscalar mesons, the Goldstone bosons connected with the spontaneous
symmetry breakdown. Their properties may be analyzed by means of an effective
Lagrangian, which describes the low energy
structure of the theory in terms of an expansion in powers of
momenta and quark masses \cite{Weinberg Physica,GL SU(3)}.

If the number of colours is taken large, the quark
loop graph which gives rise to the anomaly in the divergence of the
singlet axial current is suppressed. In the limit $N_c\!\rightarrow\!\infty$,
QCD thus acquires an additional
U(1) symmetry, whose spontaneous breakdown gives rise to a ninth Goldstone
boson, the $\eta'$ \cite{Large Nc,Coleman Witten,Minkowski}.
The implications for the effective Lagrangian are extensively discussed in
the literature \cite{Leff U(3)} and the leading terms in the
expansion in powers of $1/N_c$ have been worked out.
The purpose of the present paper is to extend this analysis to first nonleading
order and to discuss the consequences for the mass spectrum of the
pseudoscalars.

The
relevant dynamical variable is
the Goldstone field $U(x)$, which lives on the quotient G/H, where G and H are
the (approximate) symmetry groups of the Hamiltonian and of the ground
state, respectively. Since I wish to study the large $N_c$ limit, G
is the group
U(3)$_\R\times$U(3)$_\L$ and $U(x)\!\!\in$U(3). The unimodular part of
the field $U(x)$ contains the degrees of freedom of the pseudoscalar
octet, while the phase $\mbox{det}\,U(x)=e^{i\phi_0(x)}$ describes the
$\eta^\prime$.

The effective Lagrangian is formed with the field $U(x)$ and its derivatives,
${\cal L}_\eff\!=\!{\cal L}_\eff (U,\partial U,\partial^2
U,\ldots)$. The low energy analysis relies on the symmetry properties of this
function, which follow from the fact that the Lagrangian of QCD
is invariant under
independent U(3) rotations of the right- and lefthanded components of the
quark fields -- except for the U(1) anomaly which ruins the conservation of
the axial singlet current and for the explicit symmetry breaking due to the
quark masses. In an arbitrary chiral basis, the quark mass term is of the form
$\qbar_\R m q_\L\!+\qbar_\L m^\dagger q_\R$. Using the projectors
$P_\R=\frac{1}{2}(1+\gamma_5)$, $P_\L=\frac{1}{2}(1-\gamma_5)$, the Dirac
operator may be written as $D=-i\gamma^\mu
D_\mu +P_\L m +P_\R m^\dagger$. The chiral rotation $q_\R\!\rightarrow \!
V_\R\,q_\R$, $q_\L\!\rightarrow \!V_\L\, q_\L$, $V_\R,V_\L\!\in\!\mbox{U(3)}$
takes this operator into $D'=VD\,\tilde{V}$, with $V=P_\R V_\R +P_\L V_\L$,
$\tilde{V}=
\gamma^0 V^\dagger\gamma^0=P_\L V_\R^{\,\dagger} +P_\R V_\L^{\,\dagger}$. While
the kinetic
term is invariant under the operation, the mass term transforms according to
$m'=V_\R mV_\L^\dagger$.

The path integral over the
quark degrees of freedom is
given by the determinant of the Dirac operator.
The singlet axial current is anomalous because
the determinant does not remain the
same under the above operation, but picks up a phase, $\mbox{det}D'=
 e^{i\alpha \nu}  \mbox{det}D$,
where
$\alpha$ is the angle of the U(1) rotation,
$e^{i\alpha}\!=\!\mbox{det}\,V_\R V_\L^\dagger$, and
\bdm \nu=\int\!\!dx \,\omega\co\;\;\;
\omega=\frac{1}{32\pi^2}G_{\mu\nu}^{\,a}\tilde{G}^{\,a\mu\nu}\fs\edm
In euclidean space, $\nu$ is the winding
number of the field configuration.
Note that I am absorbing the coupling constant
in the gluon field, $D_\mu=\partial_\mu-iG_\mu$.

The change in the Dirac determinant in
effect adds
a term to the Lagrangian proportional to $\omega$. So, if a term of this form
is allowed for to start with,
$\Lbar_\QCD\!=\!{\cal L}_\QCD-\theta\,\omega$, the form of
the Lagrangian remains the same under the full set of chiral
transformations, but both the quark mass matrix and
the vacuum angle $\theta$ undergo a change, $m\!\rightarrow\!V_\R m
V_\L^\dagger$, $\theta\!\rightarrow\theta-\alpha$.

This property of QCD may readily be formulated at
the level of the effective theory. The presence of an additional coupling
constant, $\theta$, also shows up in the effective Lagrangian, ${\cal
L}_\eff\!\rightarrow\! \Lbar_\eff\!=\!
\Lbar_\eff(U,\partial U,\ldots\,,m,\theta)$. The expression must have the
property that the transformation
\bdm U^\prime\!=\!V_\R U V_\L^\dagger\co\;\;\; m^\prime\!=\! V_\R
\,m V_\L^\dagger\co\;\;\; \theta^\prime\!=\!\theta-\alpha\edm
leaves it invariant,
$\Lbar_\eff(U^\prime,\partial U',\ldots\,,m^\prime,\theta^\prime)\!=\!
\Lbar_\eff(U,\partial U,\ldots\,,m,\theta)$.\footnote{More precisely, the
corresponding action
is invariant -- the effective Lagrangian changes by a total derivative. The
phenomenon gives rise to the Wess-Zumino-term.
Since the corresponding vertices
involve five or more meson fields, they do not contribute to the
mass formulae and I therefore disregard
the complication.}

Since the phase of the determinant $e^{i\phi_0}\!=\mbox{det}\,U$
transforms according to $\phi_0^\prime\!=\!\phi_0+\alpha$, the
combination $\phi_0+\theta$ remains
invariant. It is convenient to replace the variable $\theta$ by this
combination, $\Lbar_\eff\!=\!\Lbar_\eff(U,\partial
U,\ldots\,,m,\phi_0+\theta)$. The above symmetry relation then implies
that the effective Lagrangian is invariant under a simultaneous rotation of
the matrices $U$ and $m$, at a fixed value of the last argument. The
expansion in powers of derivatives and quark mass matrices yields a
sequence of invariants which can be formed with $m,U,\partial U,\ldots\,$.
There are two differences compared to the corresponding series of
terms which occur in the standard framework, where the $\eta'$ does
not represent a dynamical variable: the determinant of $U$ differs
from one and the coefficients of the expansion are not constants,
but functions of the variable $\phi_0+\theta$. Up to and including terms
with two derivatives or one factor of $m$, the most general expression
consistent with the symmetries of the QCD Hamiltonian reads\footnote{
Note that the trace
$\langle\partial_\mu U U^\dagger\rangle$ coincides with
$i\,\partial_\mu\phi_0$.} \cite{GL SU(3),Leff U(3)}
\be\label{b1} \Lbar_\eff =-V_0+V_1\,\langle\partial_\mu
U^\dagger\partial^\mu U\rangle+V_2\,
\langle U^\dagger m\rangle + V_2^\star\,\langle m^\dagger U\rangle+
V_3\,\partial_\mu\phi_0\partial^\mu\phi_0+\ldots\co\ee
with $V_n\!=\!V_n(\theta+\phi_0)$.
The expression is valid to all orders in $1/N_c$. The symmetry
does not constrain the form of the coefficient functions, except that
$V_0,V_1,V_3$ must be real and even under parity, $V_n(-x)\!=\!V_n(x)$, while
$V_2$ obeys $V_2(-x)\!=\!V_2(x)^\star$.

In the large $N_c$ limit, the coefficients reduce to constants. The
reason is that, in this limit, the dependence of the
various Green functions and matrix elements on $\theta$ is suppressed
\cite{Large Nc}: Compared to the leading contribution of the
QCD Lagrangian,
$-G_{\mu\nu}^{\,a}G^{\,a\mu\nu}/4g^2$, the term $-\theta\,
G_{\mu\nu}^{\,a}\tilde{G}^{\,a\mu\nu}/32\pi^2$ represents a small perturbation
of order $g^2\theta\!\sim\!\theta/N_c$.
The vacuum energy density, e.g., is of the form
$\epsilon\!=\!N_c^2 g_0(\theta/N_c)+N_c g_1(\theta/N_c)+\ldots$
The $\theta$-dependence of the
coefficients occurring
in the effective Lagrangian is similar
\cite{GL SU(3),Leff U(3)}:
\bdm V_n(\phi_0+\theta)=(N_c)^{p_n}
\{V_n^0(\tilde{\theta})+
N_c^{-1}V_n^1(\tilde{\theta})+\ldots
\}\co\;\;\;\tilde{\theta}=(\phi_0+\theta)/N_c\co\edm
with $p_0\!=\!2,p_1\!=\!p_2\!=\!1,p_3\!=\!0$. In fact,
the leading term in the expansion
of $V_0$ coincides with the gluonic contribution to the vacuum energy,
$V_0^0(x)=g_0(x)$.

The above representation shows that
the Taylor series of the coefficients with respect to $\phi_0+\theta$
is accompanied by powers of $1/N_c$, so that only the first
few terms are needed to a given order of the $1/N_c$ expansion.
The Taylor series of $V_0$, e.g., yields
$V_0(\phi_0+\theta)\!=\!\epsilon_0+\frac{1}{2}\tau(\phi_0+\theta)^2+O(1/N_c)$.
The first term $\epsilon_0\!=\!O(N_c^{\,2})$ is an irrelevant cosmological
constant, while $\tau\!=\!O(N_c^{\,0})$ represents
the second derivative of the gluonic contribution to the vacuum energy with
respect to $\theta$ and is referred to as the topological susceptibility.

In the present context,
the dependence on $\theta$ is of interest only to incorporate
the anomalous Ward identity and to control
the expansion in powers of $1/N_c$. The vacuum angle has now served its
purpose and can be dismissed, setting $\theta\!=\!0$. Also, I now choose the
chiral basis in such a manner that the quark mass matrix is diagonal and real.

The preceding analysis involves a simultaneous expansion in powers
of $1/N_c$, powers of momenta $p$ and powers of the quark mass matrix $m$.
It is convenient to order this triple series
by counting the three expansion parameters as small quantities of order
$1/N_c\!=\!O(\delta)$,
$p\!=\!O(\sqrt{\delta})$ and $m\!=\!O(\delta)$, respectively. Disregarding the
cosmological constant mentioned above, the expansion then
takes the form ${\cal L}_\eff={\cal L}_\eff^{(0)} + {\cal
L}_\eff^{(1)}+\ldots\,$, where the first term is of order one,while the
second collects the corrections of $O(\delta)$. Setting
$V_1(0)\!=\!\frac{1}{4}F^2$, $V_2(0)\!=\!\frac{1}{2}F^2B$, the explicit
expression
for the leading term reads
\bdm {\cal L}_\eff^{(0)}= \mbox{$\frac{1}{4}$}F^2
\langle\partial_\mu U^\dagger\partial^\mu U\rangle
+\mbox{$\frac{1}{2}$} F^2 B\langle m (U  +
U^\dagger)\rangle-\mbox{$\frac{1}{2}$}\tau\phi_0^2 \fs\edm
The first two terms are familiar from the standard effective Lagrangian. They
involve the pion decay constant
$F\!=\!O(\sqrt{N_c})$ and the constant $B\!=\!O(N_c^{\,0})$, which is related
to the quark condensate. The third term is characteristic of the extension from
SU(3)$_\R\times$SU(3)$_\L$ to U(3)$_\R\times$U(3)$_\L$. It equips the $\eta'$
with a mass proportional to the square root of the topological susceptibility.

The following calculation also accounts for the corrections of first
nonleading order. In addition to the terms arising from the Taylor
coefficients $V_2'(0)\!\equiv\! 2 i B K_1$ and
$V_3(0)\!\equiv\!K_2$, which both are of $O(N_c^{\,0})$, these
also involve contributions
from higher orders of the derivative expansion, omitted in eq.(\ref{b1}).
Their structure is known from the standard
framework, where the corresponding effective coupling
constants are denoted by $L_1,\ldots\,,L_8 $. The first three of these do not
play any role in the following, because they multiply invariants of the type
$(\partial U)^4$ and do therefore not
contribute to the masses of the pseudoscalars. The term $L_4
\langle\partial_\mu U^\dagger\partial^\mu
U\rangle\langle m U^\dagger+mU\rangle$ contains two traces.
Contributions of this structure can only arise
from graphs with two or more quark loops: The term violates
the Okubo-Iizuka-Zweig rule and is suppressed by one power of $1/N_c$. The
same applies to the contributions proportional to $L_6$ and $L_7$. The
coupling constants $L_5$ and $L_8$, however, are of $O(N_c)$ and do contribute
to the masses at first nonleading order of the above expansion.
Absorbing the constant $ B$ in the quark mass matrix with $\chi\!\equiv\! 2 B
m$, the effective Lagrangian becomes
\bea
{\cal L}_\eff\al=\al \mbox{$\frac{1}{4}$}F^2
\langle\partial_\mu U^\dagger\partial^\mu U+
\chi( U  + U^\dagger)
\rangle-\mbox{$\frac{1}{2}$}\tau\phi_0^2 \\
\al\al+ L_5\langle\partial U^\dagger \partial U(\chi U +U^\dagger\chi)
\rangle + L_8\langle U \chi U\chi +\chi
U^\dagger\chi U^\dagger \rangle \no \al\al
+ i K_1 \phi_0\langle \chi(U^\dagger- U)\rangle
+K_2\partial_\mu\phi_0\partial^\mu\phi_0+O(\delta^2) \fs\nonumber
 \eea
The first line contains the leading contributions of order $N_c \,p^2,N_c \,m$
and $N_c^{\,0}$, respectively. Their relative size depends on the relative
magnitude of the three expansion parameters $1/N_c,p,m$. The second line
contains the corrections of order $N_c\,p^2\,m$ and $N_c\, m^2$, while the
third one accounts for those of order $N_c^{\,0}\,m$ and $N_c^{\,0}\,p^2$.

The mass spectrum is obtained by setting $U\!=\!\exp\,i\varphi/F$ and working
out the terms quadratic in the matrix field $\varphi$. In this notation,
$\phi_0$ is given by the trace $\langle\varphi\rangle/F$, so
that the quadratic terms are
\bea\label{b2} {\cal
L}_\eff\al=\al\mbox{$\frac{1}{4}$}\langle\partial_\mu\varphi\,\partial^\mu
\varphi\rangle
+2L_5F^{-2}
\langle\chi\,\partial_\mu\varphi\,\partial^\mu\varphi\rangle
+K_2F^{-2}\langle\partial_\mu\varphi\rangle\langle\partial^\mu
\varphi\rangle\\\al-\al
\mbox{$\frac{1}{4}$}\langle\chi\,\varphi^2\rangle -\mbox{$\frac{1}{2}$}
\tau F^{-2}\langle\varphi\rangle^2
-2L_8F^{-2}\langle\chi\varphi\chi\varphi+\chi^2\varphi^2\rangle
+2 K_1F^{-2}\langle\varphi\rangle\langle\chi\varphi\rangle
\fs\nonumber\eea
For those fields which carry electric charge or strangeness, this expression is
diagonal and yields
\be M_{\pi^+}^2= (m_u+m_d)B\{1 +8(m_u+m_d)(2L_8-L_5)BF^{-2}\}\co\ee
and analogously for $M_{K^+}^2,M_{K^0}^2$. These relations agree with those of
chiral perturbation theory \cite{GL SU(3)}, with two simplifications: (i) The
coupling constants $L_4$ and $L_6$ do not occur here, because they are
suppressed by one power of $N_c$. (ii) For the same reason, the chiral
logarithms generated
by the one loop graphs are absent. Since these graphs are inversely
proportional to $F^2$, they only show up at the next order of the expansion
under consideration.

The above mass formulae imply that the
corrections in the two ratios\footnote{The quantity $\hat{m}$
denotes the mean mass of $u$ and $d$,
$\hat{m}\equiv\frac{1}{2}(m_u+m_d)$}
\bea \frac{M_K^2}{M_\pi^2}\al=\al\frac{m_s+\hat{m}}{2
\hat{m}}\{1+\Delta_M\} \no
\frac{M_{K^0}^2-M_{K^+}^2}{M_K^2-M_\pi^2}\al=\al\frac{m_d-m_u}{m_s-\hat{m}}
\{1+\Delta_M\}\nonumber\eea
are the same. Eliminating the quark
masses
in favour of $M_\pi,M_K$, the explicit expression for the correction becomes
\be\label{DeltaM}
\Delta_M=\frac{8}{F^2}(M_K^2-M_\pi^2)(2L_8-L_5)\fs
\ee
In the double ratio
\bdm Q^2\equiv \frac{M_K^2}{M_\pi^2}\;\frac{M_K^2-M_\pi^2}{
M_{K^0}^2-M_{K^+}^2}\co\edm the correction
drops out. Accordingly, the corresponding ratio of quark masses is
determined by the masses of the pseudoscalars
\be\label{ellipse} \frac{m_s^2-\frac{1}{4}(m_u+m_d)^2}{m_d^2-m_u^2}=Q^2\co\ee
up to and including first order corrections \cite{GL SU(3)}.
Note that I have disregarded the
electromagnetic interaction. Evaluating the corresponding self energies
with the Dashen theorem \cite{Dashen},
one finds $Q\simeq 24$. The theorem holds only in the chiral limit.
The corrections of higher order are discussed in \cite{DHW Bijnens,Urech}. The
corresponding uncertainty in the value of $Q$ is of order 10 \%.

In the neutral sector,
the contributions from $L_5$ and $K_2$ introduce off-diagonal elements into the
kinetic term. These are removed with the change of variables
$\varphi\!\rightarrow\varphi
-4\{m, \varphi\}L_5B F^{-2} - 2\langle\varphi\rangle K_2
F^{-2}$. The operation reduces the first line in eq.(\ref{b2}) to the term
$\frac{1}{4}\langle\partial_\mu\varphi\partial^\mu\varphi\rangle$ and
replaces the coefficients of the quadratic form in the second line by
\bdm\overline{\tau}=
\tau\,(1-12K_2F^{-2})\co\;\;\;
{\overline{\rule[0.75em]{0.5em}{0em}}\hspace{-0.65em}L}_8=L_8
-\mbox{$\frac{1}{2}$}L_5 \co\;\;\;\Kbar_1=
K_1+\mbox{$\frac{1}{2}$}K_2+2 L_5\tau F^{-2}\fs\edm
The masses of the neutral particles are
thus obtained by diagonalizing this form.
The mixing angles between the
$\pi^0$ and $\eta,\eta^\prime$ are proportional to the isospin breaking mass
difference
$m_d\!-\!m_u$. These angles are small, but play a crucial role e.g. for
the transition $\eta\rightarrow3\pi$. In the masses of the neutral particles,
however, isospin
breaking only generates contributions of order $(m_d-m_u)^2$, which are
negligibly small. Disregarding these, the $\pi^0$ is
degenerate with $\pi^\pm$ and the $\eta$ only mixes with the $\eta'$.
Setting
$\varphi=\varphi_8\lambda_8+\varphi_9\sqrt{\rule[0.3em]{0em}{0.2em}}
\mbox{$\frac{2}{3}$}$,
the quadratic part of the Lagrangian becomes
\bdm{\cal
L}_\eff=\mbox{$\frac{1}{2}$} (\partial_\mu\varphi_8\partial^\mu\varphi_8
+\partial_\mu\varphi_9\partial^\mu\varphi_9)-
\mbox{$\frac{1}{2}$}(m_1^2\varphi_8^2
-2\sigma_1\varphi_8\varphi_9+M_1^2\varphi_9^2)
\fs\edm
The mass formulae for the charged particles may be used to eliminate the
quark masses in favour of
$M_K^2,M_\pi^2$. The coefficients then take the form
\bea
 m_1^2\al=\al\mbox{$\frac{1}{3}$}(4M_K^2\!-\!M_\pi^2) +
\mbox{$\frac{4}{3}$}(M_K^2\!-\!M_\pi^2)\Delta_M\no
\sigma_1\al=\al\mbox{$\frac{2}{3}$}\sqrt{2}(M_K^2\!-\!M_\pi^2)
\{1+\Delta_M-\Delta_N\}\no
M_1^2\al=\al 6\frac{\overline{\tau}}{F^2}+
\mbox{$\frac{1}{3}$}(2M_K^2\!+\!M_\pi^2)
(1\!-\!2\Delta_N)+\mbox{$\frac{2}{3}$}(M_K^2\!-\!M_\pi^2)\Delta_M\fs
\nonumber\eea
Remarkably, $m_1^2$ only involves the same correction
$\Delta_M$ which also occurs in the mass formulae for $\pi$ and $K$. The
term
$\Delta_N=12\Kbar_1/F^2$,
which describes the Zweig rule violating contributions of order $1/N_c$, only
affects the quantities $\sigma_1$ and $M_1^2$.

The eigenvalues $M^2\!=\!(M_\eta^2,M_{\eta'}^2)$ obey
$(m_1^2-M^2)(M_1^2-M^2)\!=\!\sigma_1^2$. Eliminating $M_1$, this yields
$(m_1^2-M_\eta^2)
(M_{\eta'}^2-m_1^2)\!=\!\sigma_1^2$ or, equivalently,
\be \label{mass
formula}M_\eta^2=m_1^2-\frac{\sigma_1^2}{M_{\eta^\prime}^2-m_1^2}\fs\ee
The relation states that the Gell-Mann-Okubo formula,
$\mbox{$M_\eta^2\!=\!\frac{1}{3}(4M_K^2\!-\!M_\pi^2)$}$,
receives two categories of SU(3) breaking corrections:
While the first is governed by the same parameter $\Delta_M$ which also
determines the corrections in the masses of the charged
particles and is accounted for in the term $m_1^2$, the second arises from
the $\eta$-$\eta^\prime$ transition matrix element of the
operator $\qbar m q$ and is proportional to $\sigma_1^2$.

The consequences of these mass formulae are the following. I first recall
that the relation (\ref{ellipse})
constrains the two ratios $x\!=\!m_u/m_d$ and $y\!=\!m_s/m_d$ to an ellipse.
Since $Q^2$ is very large,
the ellipse is well approximated by
$\mbox{$x^2+y^2/Q^2\!=\!1$}$, i.e. the center is at
the origin and $Q$ represents the large semiaxis, while the
small one is equal to $1$.
To determine the individual ratios $x,y$ one needs to know the quantity
$\Delta_M$, which describes the strength of SU(3) breaking in the mass
formulae and involves the two effective coupling
constants $L_5,L_8$. The former is known from the
observed asymmetry in the decay constants $F_\pi,F_K$, but, as pointed out by
Kaplan and Manohar \cite{Kaplan Manohar}, $L_8$
cannot be determined on purely phenomenologigal grounds.
The scattering of the chiral perturbation theory results for the mass ratios
$m_u/m_d$
and $m_s/m_d$ encountered in the literature \cite{GL74/75}--\cite{Wyler}
originates in this problem:
Treating $L_8$
as a free parameter, one may obtain any value for $\Delta_M$ and thus reach any
point on the ellipse. In particular,
the possibility that the mass of the
lightest quark might vanish
is widely discussed in the literature,
because
this would remove the strong CP problem \cite{Peccei}. This possibility
corresponds to $m_u/m_d\!=\!0,\,m_s/m_d\!=\!Q$ and requires that the
correction is large and negative,
$\Delta_M\!=\!M_K^2/M_\pi^2/(Q+\frac{1}{2})-1\simeq
-0.45$.

I now wish to show that the
framework specified above leads to a lower bound on $\Delta_M$ which
requires $m_u$ to be different from zero.
The bound arises because
the relation (\ref{mass formula}) only admits a solution for
$M_\eta^2\!<\!m_1^2$, or \be \label{first inequality}
\Delta_M>-\frac{4M_K^2-3M_\eta^2-M_\pi^2}{4(M_K^2-M_\pi^2)}=-0.07\fs\ee
The inequality is an immediate consequence of the
fact that $\eta$-$\eta^\prime$ mixing leads to a repulsion of
the two levels.
Admittedly, the
hypothesis that the first two terms of the $1/N_c$ expansion yield a decent
approximation for the theory of physical interest, $N_c\!=\!3$, goes beyond
solid phenomenology. This hypothesis,
however, represents the only coherent
explanation of the fact that the
Okubo-Zweig-Iizuka-rule holds to a good approximation
and I see no reason to doubt its reliability in the present
case.

The large negative value of $\Delta_M$ required if $m_u$ were to vanish
violates the above bound.
The $1/N_c$ expansion thus corroborates the conclusion that
$m_u$ is different from zero. As is well-known,
the quantity $e^{i\theta}\det m$ then represents a physically significant
parameter of the QCD Lagrangian. Why the mechanism which generates $m$ and
$\theta$ happens to be such that, to a high degree of accuracy,
this quantity
is real, represents a well-known puzzle \cite{Peccei}. It cannot be
solved the easy way, with $m_u\!=\!0$.

A similar inequality also holds
for the $1/N_c$ correction,
\bdm
\Delta_N>3\frac{(\sqrt{3}+1)M_\eta^2-(\sqrt{3}-1)M_{\eta^\prime}^2
-2 M_\pi^2}{8(M_K^2-M_\pi^2)}=0.18\fs\edm
This indicates that Zweig rule violating
effects are not entirely negligible.

As it stands, the mass formula (\ref{mass formula})
only determines the magnitude of the SU(3) asymmetry
$\Delta_M\!\sim\!(m_s\!-\!\hat{m})$ if the size of the Zweig rule violation
$\Delta_N\!\sim\!1/N_c$ is known and vice versa.
One may, e.g., choose a value for $\Delta_N$, such that
$\Delta_M$ saturates the inequality
(\ref{first inequality}). This is the case, however, only if $\sigma_1$
vanishes, i.e. if
$1+\Delta_N-\Delta_M\!=\!0$. In other words, the corrections would have
to cancel the leading term. It is clear that
in such a situation, the above formulae
are meaningless.
A coherent picture only results if both
$|\Delta_M|$ and $|\Delta_N|$ are small compared to unity.
The mass formula (\ref{mass formula}) shows that a negative value of
the SU(3) asymmetry $\Delta_M$ requires exorbitant $1/N_c$ corrections.
Even $\Delta_M\!=\!0$ calls for large
Zweig rule violations,
$\Delta_N\simeq\frac{1}{2}$.
The term $(1\!+\!\Delta_M\!-\!\Delta_N)^2$, which
corrects the contribution generated by
the singlet-octet transitions for effects of first nonleading order, would then
instead modify this contribution by a factor of 4
if evaluated as it stands, eliminating it
altogether if expanded to first order in the "corrections".
The condition
\be \Delta_M\!>\!0\ee
thus represents a generous lower bound for
the region where a truncated $1/N_c$ expansion leads to meaningful results.
It states
that the current algebra formula, which relates the quark mass ratio
$m_s/\hat{m}$ to the meson mass ratio $M_K^2/M_\pi^2$ represents an upper
limit, $m_s/\hat{m}\!<\!2M_K^2/M_\pi^2-1$. The corresponding bounds on the
two ratios $m_u/m_d$ and $m_s/m_d$ depend on the value
of $Q$,
\bea \frac{m_u}{m_d}\al>\al\frac{M_\pi^4Q^2-M_K^4+M_K^2M_\pi^2}
{M_\pi^4Q^2+M_K^4-M_K^2M_\pi^2}\co\\
 \frac{m_s}{m_d}\al<\al\frac{M_\pi^2(2M_K^2-M_\pi^2)Q^2}
{M_\pi^4Q^2+M_K^4-M_K^2M_\pi^2}\fs\eea
If $Q$ is taken from
the Dashen theorem, these
relations state that the Weinberg ratios \cite{Weinberg Rabi} correspond to
the limiting case where the bounds are saturated,
$m_u/m_d\!>\!0.55,m_s/m_d\!<\!20.1$.
Lowering the value of $Q$ to $Q\!=\! 22$, the bound on
$m_u/m_d$ becomes slightly weaker, $m_u/m_d\!>\!0.48$, while the one for
$m_s/m_d$ decreases to $m_s/m_d\!<\!19.2$.
The estimate $m_u/m_d\!=\!0.3\pm 0.1$
obtained in ref.\cite{DHW PRL,Wyler} from a multipole analysis of the
transitions $\psi^\prime\rightarrow\psi\pi^0$,
$\psi^\prime\rightarrow\psi\eta$ is outside this range and is thus not
consistent with the above arguments (see also ref.\cite{Luty Sundrum}).

To demonstrate that the
observed mass pattern is perfectly consistent with
the hypothesis that the corrections of order $1/N_c$ as well as those of
order $m$ are small, I note
that the
contribution from the second term in eq.(\ref{mass formula})
is small, because it is suppressed
by a factor of order $M_\eta^2/M_{\eta^\prime}^2$. The corrections to that
term are reduced by the same factor. The only new contribution in the second
order formula which does not get suppressed
is the one responsible for the difference between $m_1^2$ and the value of
$M_\eta^2$ which follows from the Gell-Mann-Okubo formula. Retaining only this
term, the second
order formula simplifies to
\bdm \Delta_M=-\frac{4M_K^2-M_\pi^2-3M_\eta^2}
{4(M_K^2-M_\pi^2)}+
\frac{2(M_K^2-M_\pi^2)}{3M_{\eta^\prime}^2+M_\pi^2-4M_K^2}
\fs\edm
Numerically, this gives
$\Delta_M\!=\!0.18$, thus requiring a breaking of SU(3)
symmetry in the mass formulae of the same order of magnitude as in the decay
constants. The corresponding value for the $1/N_c$ correction is also
quite small, $\Delta_N\!=\!0.24$. I repeat, however, that the above framework
does not predict the values of the two corrections individually, but only
correlates them. In particular,
the truncated large $N_c$ expansion is also consistent with
a somewhat smaller breaking of SU(3) symmetry and a
correspondingly larger violation of the Zweig rule.

\end{document}